\documentclass[iop,apjl]{emulateapj}
\usepackage{amsmath,amssymb,amstext}

\usepackage[backref,breaklinks,colorlinks,citecolor=blue,linkcolor=magenta]{hyperref}

\usepackage[all]{hypcap} 

\usepackage{aas_macros}
\usepackage{natbib}
\shorttitle{Ophiuchus Stream Fanning}
\shortauthors{Sesar et al.}

\begin{document}

\title{Evidence of Fanning in the Ophiuchus Stream}

\author{Branimir Sesar\altaffilmark{1}}
\author{Adrian M.~Price-Whelan\altaffilmark{2}}
\author{Judith G.~Cohen\altaffilmark{3}}
\author{Hans-Walter Rix\altaffilmark{1}}
\author{Sarah Pearson\altaffilmark{2}}
\author{Kathryn V.~Johnston\altaffilmark{2}}
\author{Edouard J.~Bernard\altaffilmark{4}}
\author{Annette M.~N.~Ferguson\altaffilmark{4}}
\author{Nicolas F.~Martin\altaffilmark{5,1}}
\author{Colin T.~Slater\altaffilmark{6}}
\author{Kenneth C.~Chambers\altaffilmark{7}}
\author{Heather Flewelling\altaffilmark{7}}
\author{Richard J.~Wainscoat\altaffilmark{7}}
\author{Christopher Waters\altaffilmark{7}}
\altaffiltext{1}{Max Planck Institute for Astronomy, K\"{o}nigstuhl 17, D-69117 Heidelberg, Germany; \email{bsesar@mpia.de}}
\altaffiltext{2}{Department of Astronomy, Columbia University, 550 West 120th Street, New York, NY 10027, USA}
\altaffiltext{3}{Division of Physics, Mathematics and Astronomy, California Institute of Technology, Pasadena, CA 91125, USA}
\altaffiltext{4}{SUPA, Institute for Astronomy, University of Edinburgh, Royal Observatory, Blackford Hill, Edinburgh EH9 3HJ, UK}
\altaffiltext{5}{Observatoire astronomique de Strasbourg, Universit\'e de Strasbourg, CNRS, UMR 7550, 11 rue de l'Universit\'e, F-67000 Strasbourg, France}
\altaffiltext{6}{Department of Astronomy, University of Michigan, 1085 S. University Ave., Ann Arbor, MI 48109, USA}
\altaffiltext{7}{Institute for Astronomy, University of Hawaii at Manoa, Honolulu, HI 96822, USA}

\begin{abstract}
The Ophiuchus stellar stream presents a dynamical puzzle: its old stellar populations ($\sim 12$ Gyr) cannot be reconciled with (1) its orbit in a simple model for the Milky Way potential and (2) its short angular extent, both of which imply that the observed stream formed within the last $<1$~Gyr. Recent theoretical work has shown that streams on chaotic orbits may abruptly fan out near their apparent ends; stars in these fans are dispersed in both position and velocity and may be difficult to associate with the stream. Here we present the first evidence of such stream-fanning in the Ophiuchus stream, traced by four blue horizontal branch (BHB) stars beyond the apparent end of the stream. These stars stand out from the background by their high velocities ($v_{\rm los} > 230$ km~s$^{-1}$) against $\sim 40$ other stars: their velocities are comparable to those of the stream, but would be exceptional if they were unrelated halo stars. Their positions and velocities are, however, inconsistent with simple extrapolation of the observed cold, high-density portion of the stream. These observations suggest that stream-fanning may be a real, observable effect and, therefore, that Ophiuchus may be on a chaotic orbit. They also show that the Ophiuchus stream is more extended and hence dynamically older than previously thought, easing the stellar population {\it vs.} dynamical age tension.
\end{abstract}


\maketitle

\section{Introduction}\label{introduction}

Cold stellar streams are remnants of Milky Way satellites or globular clusters that were disrupted by tidal forces and stretched into filaments as they have orbited in the Galaxy's potential (e.g., GD-1, \citealt{gd06}; Orphan stream, \citealt{gri06,bel07a}). As ensembles of stars on closely related orbits, their position, kinematics, and morphology are sensitive to the underlying potential and thus can be used to constrain it \citep{kop10,new10,bon14,apw14}. The angular extent of cold stellar streams, in conjunction with their velocity dispersion and orbit, had been taken as an estimator for their dynamical disruption ages. 

The recently found cold (velocity dispersion $\sigma_v\lessapprox 1$~km~s$^{-1}$) Ophiuchus stream \citep{ber14b}, presents a puzzle in this respect: its stellar population is old ($\sim 12$~Gyr), yet it is on an orbit with an orbital period $P\approx350$~Myr and it is apparently only 1.6~kpc long \citep[hereafter S15]{ses15}. To maintain such a short length, the observed debris should have been disrupted only within the last 250 million years.

Recent theoretical work on streams \citep{pea15} found that cold $N$-body streams simulated in a triaxial potential can exhibit a rather abrupt spreading in position and velocity space at the seeming ends of the cold stream, which they named {\em stream-fanning} (see their Figure 4). This fanning may arise if the progenitor is on a chaotic orbit, common in triaxial or otherwise complex gravitational potentials (bottom panels of Figure 11 of \citealt{far15}; Figure 7 of \citealt{apw16}); or it may arise through interactions with dark matter subhalos (bottom left panel of Figure 3 of \citealt{bon14}). Hence, stream-fanning--if observable--could become a new tool for studying these dynamical phenomena. In a first application, \citet{pea15} used the {\em absence} of fanning in the known long and cold stream of the globular cluster Palomar 5 (Pal 5) \citep{ode01} to rule out the \citet{lm10} (triaxial) dark matter halo potential within $r\lesssim25$ kpc.

Stream-fanning may turn out to be the most plausible explanation for the stellar population {\it vs.} age tension in the Ophiuchus stream. One possible solution to this discrepancy could be that the stream is in reality much longer (i.e., has been undergoing disruption for a much longer time), but is difficult to detect in imaging alone due to very low surface brightness caused by stream-fanning. Stream-fanning in the case of the Ophiuchus stream should perhaps even be expected, as near its pericenter ($\sim 3$~kpc) the triaxiality and time-dependence of the Milky Way's stellar bar must play a role \citep{apw16} 

In this paper we present a search for, and discovery of, Ophiuchus stream members ``fanned-out'' well beyond its clearly limited narrow and kinematically cold extent (see \autoref{fig1}). The selection and spectroscopic followup of candidate stream members are described in \autoref{data}. To identify stream members we have used radial velocity measurements, since the stream's heliocentric velocity of $\sim290$ km~s$^{-1}$ is very different from the typical velocities of halo stars ($\sim100$ km~s$^{-1}$ wide Gaussian centered at zero, \citealt{xue08}). The velocity measurements and the newly identified stream members are reported in \autoref{results}, and the results are discussed in \autoref{discussion}.

\section{Data}\label{data}

\begin{figure*}
\plotone{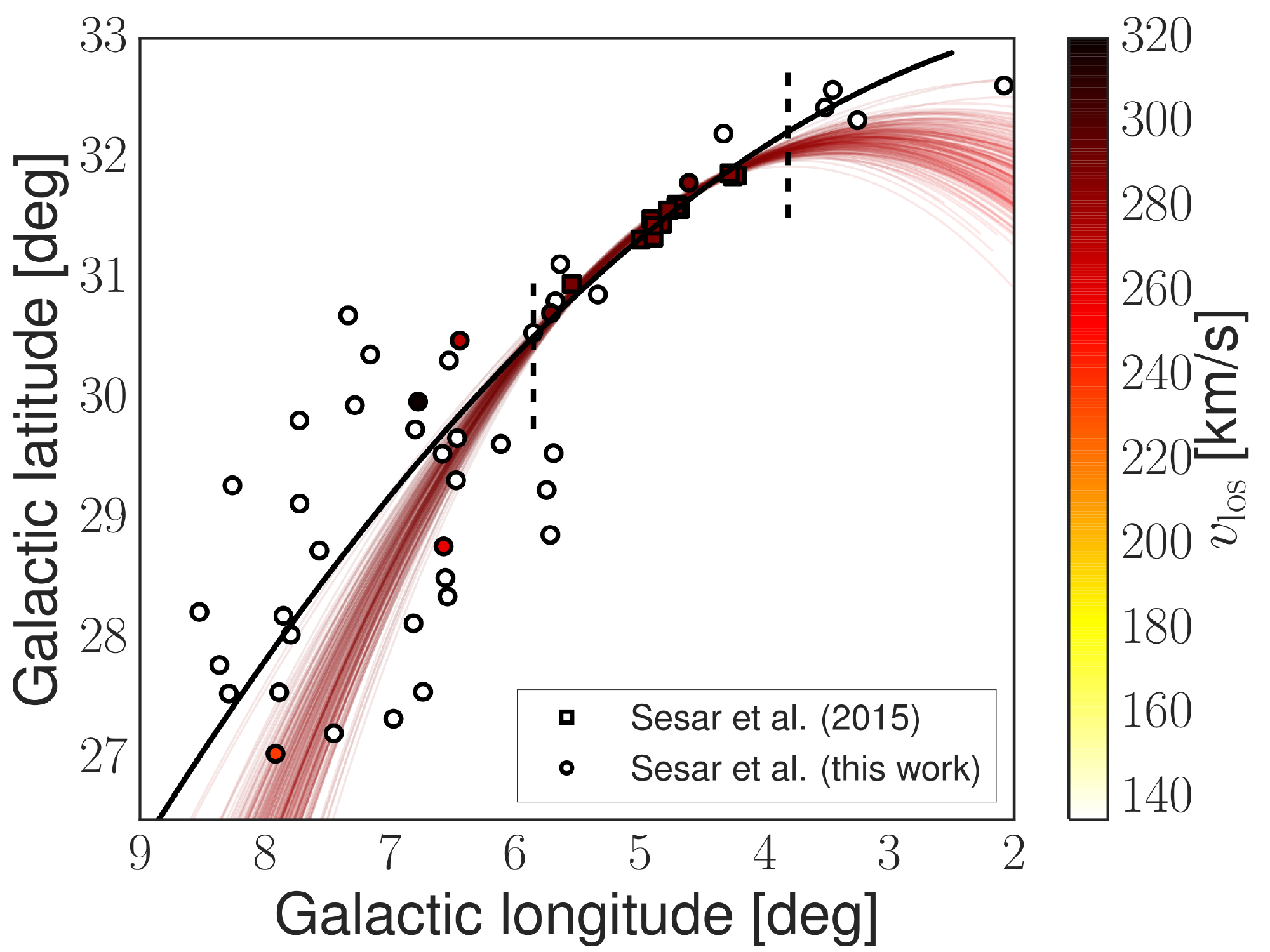}
\caption{
Positions of stars followed up in this work and by S15. The solid line shows the {\em extrapolation} of the most probable model for the ridge line of the stream, and vertical dashed lines show the apparent extent of the stream (as inferred from PS1 photometric data only, see Section 3.4 and Figure 7 of S15). The thin lines illustrate the uncertainty in the orbit of the Ophiuchus stream, and the gradient of their color indicates the predicted $v_{\rm los}$ (the predicted $v_{\rm los}$ ranges from 260 to 290 km~s$^{-1}$). Similarly, the color of symbols indicates the $v_{\rm los}$ of stars. A histogram of velocities is shown in \autoref{fig2}. Note that the majority of stars have $v_{\rm los}$ {\em smaller} than 140 km~s$^{-1}$ (open/white circles). In \autoref{results}, we argue that the four stars with $v_{\rm los} > 230$ km~s$^{-1}$ ($v_{\rm gsr} > 280$ km~s$^{-1}$) and galactic longitude $\ell>6\arcdeg$ are new Ophiuchus members that trace the fanning of the stream.
\label{fig1}}
\end{figure*}

\begin{figure}
\plotone{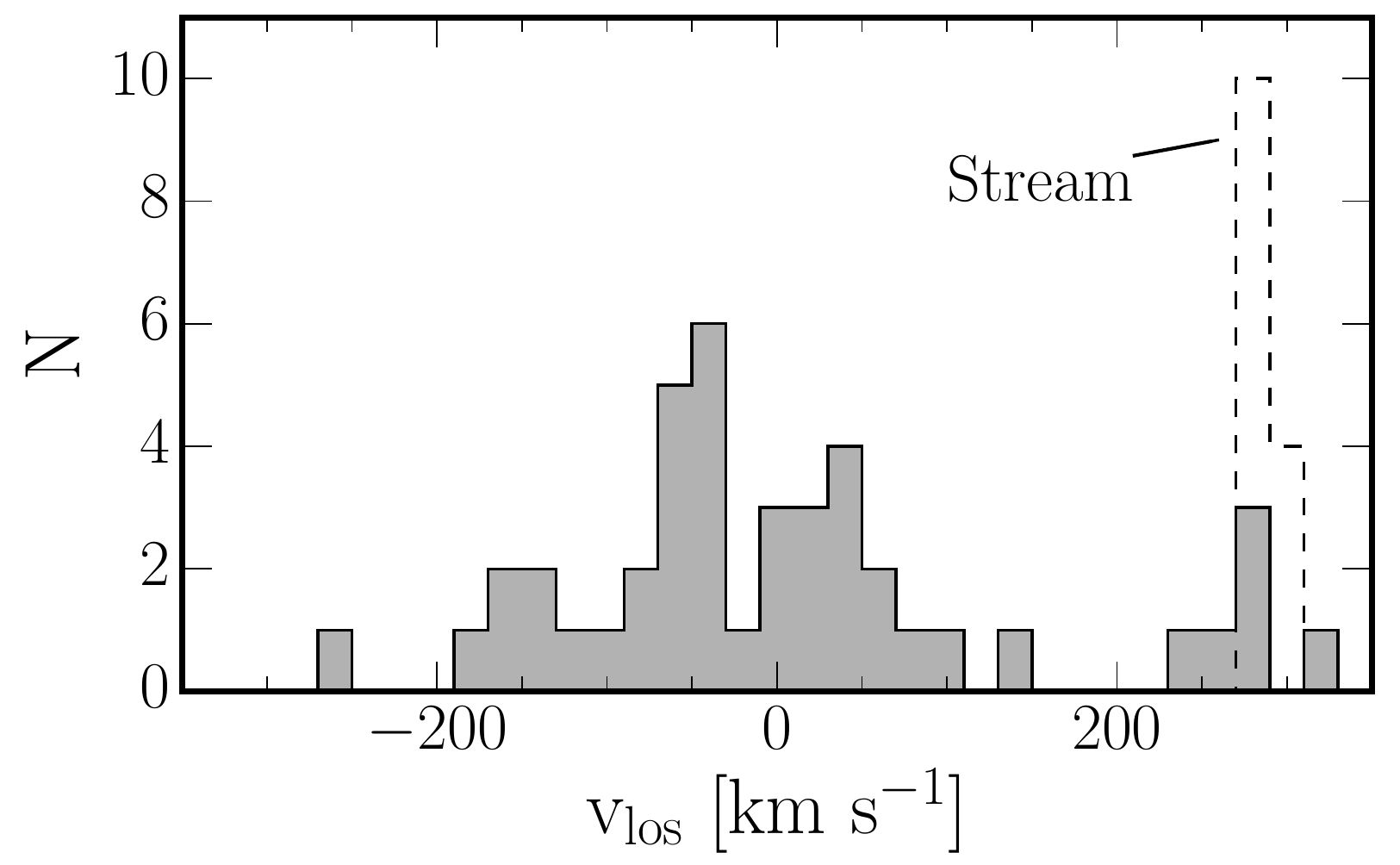}
\caption{
A histogram of heliocentric $v_{\rm los}$ velocities shown in \autoref{fig1}. The distributions of stars followed-up in this work and by S15 are shown with solid and dashed histograms, respectively. The width of bins is 20 km~s$^{-1}$.
\label{fig2}}
\end{figure}

In \autoref{fig1}, we show the spatial distribution of candidate blue horizontal branch (BHB) stars selected for spectroscopic followup, and in \autoref{table1} we list their positions. A histogram of heliocentric velocities of stars from \autoref{fig1} is shown in \autoref{fig2}.

The targets selected for spectroscopic followup have dereddened\footnote{Using the \citet{SFD98} dust map and extinction coefficients for PS1 bands by \citet{sf11}.} $g_{\rm P1}-i_{\rm P1}$ color and $g_{\rm P1}$-band magnitude consistent with the horizontal branch of the Ophiuchus stream (within 0.2 mag in magnitude or 0.1 mag in color, see Figure 6 of S15). When selecting these targets, we took into account the fact that the distance modulus of the stream changes as a function of galactic longitude $\ell$
\begin{equation}
DM(\ell) = 14.58 - 0.2(\ell - \ell_0)\label{stream_dm},
\end{equation}
where $\ell_0=5\arcdeg$ (S15). In addition, we required that the targets have dereddened color $g_{\rm P1}-i_{\rm P1} < 0$ (i.e., they need to be on the blue side of the horizontal branch) and be within $2\arcdeg$ of the extrapolated ridge line of the stream (solid line in \autoref{fig1}), which is defined as $b(\ell) = -0.15(\ell - \ell_0)^2 -0.8(\ell - \ell_0) + 31.37$ (S15). For comparison, the $2\arcdeg$ selection band around the extrapolated ridge line of stream is about eight times wider than the full-width-at-half-maximum of the stream, which is $14\arcmin$ \citep{ber14b}. 

\capstartfalse
\begin{deluxetable*}{lrrrrrrrlll}
\tabletypesize{\scriptsize}
\setlength{\tabcolsep}{0.02in}
\tablecolumns{11}
\tablewidth{0pc}
\tablecaption{Positions and Velocities of Observed Candidate BHB Stars\label{table1}}
\tablehead{
\colhead{Name} & \colhead{R.A.} & \colhead{Decl.} & \colhead{$v_{los}$} & \colhead{$g_{P1}^a$} & \colhead{$i_{P1}^a$} & \colhead{$c_{\rm \gamma}^b$} & \colhead{$b_{\rm \gamma}^b$} & \colhead{Instrument} & \colhead{BHB$^c$} & \colhead{Member} \\
\colhead{ } & \colhead{(deg)} & \colhead{(deg)} & \colhead{(km s$^{-1}$)} & \colhead{(mag)} & \colhead{(mag)} & \colhead{ } & \colhead{(\AA)} & \colhead{ } & \colhead{ } & \colhead{ }
}
\startdata
cand9  & 243.19618 & -6.72622 & $289.9 \pm 2.0$ & 15.58 & 15.58 & $1.0\pm0.1$ & $8.9\pm0.5$ & DEI+TWN & Y & Y \\
cand14 & 241.74958 & -6.81383 & $287.2 \pm 2.6$ & 15.91 & 16.11 & n/a & n/a   & DEI & n/a &  Y \\
cand15 & 243.75929 & -6.36006 & $271.4 \pm 2.4$ & 16.09 & 16.19 & $1.1\pm0.1$ & $8.0\pm0.5$ & DEI+TWN & Y & Y \\
cand26 & 244.34367 & -6.43328 & $318.2 \pm 2.3$ & 16.65 & 16.66 & $1.1\pm0.2$ & $8.3\pm0.8$ & DEI+TWN & Y & Y\\
cand49 & 247.32333 & -7.34552 & $236.6 \pm 5.0$ & 15.90 & 16.00 & $1.2\pm0.1$ & $8.4\pm0.2$ & TWN & Y & Y \\
cand54 & 245.21113 & -7.29210 & $258.6 \pm 4.0$ & 15.57 & 15.31 & $0.8\pm0.1$ & $8.2\pm0.3$ & TWN & Y & Y \\
cand8  & 243.40184 & -6.72716 & $ 10.5 \pm 2.8$ & 15.67 & 15.60 & n/a & n/a & DEI & n/a & N \\
cand10 & 240.53022 & -7.10044 & $-95.6 \pm 2.7$ & 15.88 & 16.07 & n/a & n/a & DEI & n/a & N
\enddata
\tablenotetext{a}{The $g_{\rm P1}$- and $i_{\rm P1}$-band magnitudes are {\em not} corrected for extinction.}
\tablenotetext{b}{Best-fit values of the S\'{e}rsic profile fit to the H$\gamma$ line, and their 1$\sigma$ uncertainties.}
\tablenotetext{c}{A flag indicating whether the best-fit S\'{e}rsic profile of the H$\gamma$ line is consistent with those of BHB stars (within measurement uncertainties). DEIMOS spectra do not include the $H\gamma$ line, and thus stars observed exclusively by DEIMOS lack $c_{\rm \gamma}$ and $b_{\rm \gamma}$ measurements, as well as the BHB classification.}
\tablecomments{\autoref{table1} is published in its entirety in the electronic edition of the Journal. A portion is shown here for guidance regarding its form and content. } 
\end{deluxetable*}
\capstarttrue

A total of 16 stars were observed using the DEIMOS spectrograph \citep{fab02} on the WMKO Keck-II 10-m telescope (project ID 2015A-C252D, PI: J.~Cohen). The remaining 27 stars were observed using the TWIN spectrograph on the Calar Alto 3.5-m telescope (project ID H15-3.5-011, PI: B.~Sesar). The DEIMOS spectra covered the region near the Balmer $H\alpha$ line with a resolution of 1.2~{\AA} (1200~G grating), while the TWIN spectra covered Balmer lines from $H\delta$ to $H\beta$ with a 1.7~{\AA} resolution (T05 grating). The spectra were extracted and calibrated using standard IRAF\footnote{\url{http://iraf.noao.edu/}} tasks. The uncertainty in the zero-point of wavelength calibration (measured using sky lines) was $\lesssim1$ km~s$^{-1}$ for DEIMOS, and $<3$ km~s$^{-1}$ for TWIN spectra.

The line of sight velocities were measured by fitting observed spectra with synthetic template spectra selected from the \citet{mun05} spectral library\footnote{\url{http://archives.pd.astro.it/2500-10500/}}. Prior to fitting, the synthetic spectra were resampled to the same \AA~per pixel scale as the observed spectrum and convolved with an appropriate Line Spread Function. The velocity obtained from the best-fit template was corrected to the barycentric system and adopted as the line of sight velocity, $v_{\rm los}$. We added in quadrature the uncertainty in the zero-point of wavelength calibration (in km~s$^{-1}$) to the velocity error from fitting. The resulting velocities and their uncertainties are listed in \autoref{table1}. For targets observed by DEIMOS and TWIN, we list the velocity obtained from higher-resolution DEIMOS spectra (the TWIN and DEIMOS velocities agree within uncertainties).

\section{Results}\label{results}

A quick examination of \autoref{table1} shows that there are 6 stars with $v_{\rm los} > 230$ km~s$^{-1}$. Two of these stars (cand9 and cand14) are within the observed extent of the stream (see \autoref{fig1}), and their $v_{\rm los}\sim288$ km~s$^{-1}$ are consistent with the $v_{\rm los}~\sim289$ km~s$^{-1}$ of the stream measured by S15. These stars are almost certainly members of the Ophiuchus stream. The remaining four stars are located east of the stream ($\ell > 6\arcdeg$), in the region where S15 no longer observe the Ophiuchus stream in the Pan-STARRS1 catalog data (see Figure 7 of S15). Following \citet[see their Section 2.1.2]{xue08}, we have fitted the S\'{e}rsic profile \citep{ser63} to their $H\gamma$ lines and have established that they are consistent with those of BHB stars (within measurement uncertainties). The best-fit $c_{\rm \gamma}$ and $b_{\rm \gamma}$ values of the S\'{e}rsic profile are listed in \autoref{table1}.

The four BHB stars have $v_{\rm los}$ ranging from 237 km~s$^{-1}$ to 318 km~s$^{-1}$. To measure how likely it is to draw stars with $v_{\rm los}>230$ km~s$^{-1}$ from the velocity distribution of halo stars, we first calculate their velocities in the Galactic standard of rest
\begin{equation}
v_{\rm gsr} = v_{\rm los} + U\cos b \cos \ell + (v_{LSR} + V)\cos b \sin \ell + W\sin b,
\end{equation}
where $\ell$ and $b$ are galactic longitude and latitude, $v_{LSR} + V = 242$ km~s$^{-1}$ \citep{bov12}, and $(U, W)=(10, 7)$ km~s$^{-1}$ \citep{sbd10}. Given this definition, the mean and standard deviation of $v_{\rm gsr}$ of the sample with $v_{\rm gsr} < 280$ km~s$^{-1}$ (i.e., excluding the above-mentioned four BHB stars), are 8 km~s$^{-1}$ and 95 km~s$^{-1}$, respectively (the median is $-1$ km~s$^{-1}$). Assuming that values of $v_{\rm gsr}$ are drawn from a zero-mean Gaussian with standard deviation $\sigma_v = 111~{\rm km}~{\rm s}^{-1}$ (as expected for halo stars; \citealt{xue08}), the probability of drawing a velocity with $v_{\rm gsr} > 280$ km~s$^{-1}$ is $\lesssim1\%$, or $\sim0.4$ stars given the size of the observed sample ($\sim40$ stars). Thus, if velocity is the {\em only} criterion for membership with the Ophiuchus stream, then at most one of the four stars with $v_{\rm gsr}>280$ km~s$^{-1}$ may be a field halo star (i.e., is not associated with the Ophiuchus stream), and at least three stars may be associated with the stream.

As another test, we have applied the color-magnitude cuts used in target selection to a mock catalog generated from the Besan\c{c}on model \citep{rob03}. The selected mock sample contains metal-poor halo giant stars out of which $\sim1$ star per 43 stars (which is the size of our sample) has $v_{\rm gsr}>280$ km~s$^{-1}$. This result agrees with the simpler analysis presented in the previous paragraph.

\section{Discussion and Conclusions}\label{discussion}

Given the results presented in the \autoref{results}, we now discuss four hypotheses related to four BHB stars with $v_{\rm gsr} > 280$ km~s$^{-1}$ (i.e., $v_{\rm los} > 230$ km~s$^{-1}$):
\begin{enumerate}
\item The four stars are random halo stars that are not associated with the Ophiuchus stream.
\item Two of the stars (cand49 and cand54) lie on the predicted orbit and are members of the Ophiuchus stream, while the other two are halo stars not associated with the Ophiuchus stream (cand15 and cand26).
\item The four stars are members of some yet unrecognized foreground or background halo substructure (e.g., another stream) that also has a high line of sight velocity as the Ophiuchus stream.
\item The four stars trace the fanning of the Ophiuchus stream.
\end{enumerate}

The first hypothesis is not very plausible, since randomly drawing a star with $v_{\rm gsr} > 280$ km~s$^{-1}$ from the velocity distribution of halo stars is not a likely event (1 in 100, see \autoref{results}), and randomly drawing four such stars is even less likely (1 in $10^8$). For the second hypothesis to be plausible, the velocity of the Ophiuchus stream would first need to decrease by $\sim55$ km~s$^{-1}$ (i.e., from the velocity of the Ophiuchus stream to the velocity of BHB star cand49), and then increase by $\sim25$ km~s$^{-1}$ between BHB stars cand49 and cand54. Such changes in the line-of-sight velocity cannot be due to epicyclic motion of stars along tidal tails \citep{kue08}, which are $\sim5$ km~s$^{-1}$ (e.g., as in the case of the Pal 5 stream; \citealt{kue15}), making it unlikely that stars cand49 and cand54 simply trace a ``conventional'' (i.e., non-fanned) Ophiuchus stream. Furthermore, the likelihood that stars cand15 and cand26 are just random halo stars with high velocities is lower than $10^{-4}$, making the second hypothesis even less plausible.

Due to the lack of spectroscopic coverage in the $6\arcdeg < \ell < 3\arcdeg$ region (i.e., between the dashed lines in \autoref{fig1}), above and below the Ophiuchus stream in the
galactic latitude direction, it is difficult to eliminate the third hypothesis and the presence of a yet unrecognized foreground or background halo substructure in the Ophiuchus region. However, if the four stars with $v_{\rm gsr} > 280$ km~s$^{-1}$ are members of this unrecognized halo substructure, we find it highly unusual and coincidental that (1) three out of four stars from this hypothetical substructure are located within $1\arcdeg$ of the Ophiuchus stream (in the galactic longitude direction), and are not more uniformly spread over the observed region, and (2) their median velocity is consistent with the velocity of the Ophiuchus stream (see \autoref{fig2}). In the end, a more complete spectroscopic coverage of the Ophiuchus region will show whether the third hypothesis is plausible or not.

Based on the current discussion, and the fact that the four stars under consideration (1) are BHB stars, (2) are at the predicted distance of the Ophiuchus stream (by selection), (3) are in the proximity of the Ophiuchus stream and spread around its predicted orbit (see \autoref{fig1}), and (4) have $v_{\rm gsr}$ that bracket the $v_{\rm gsr}$ of the Ophiuchus stream, we conclude that they are members of the Ophiuchus stream and that they trace the fanning of the stream.

While the velocities of these stars are not consistent with the velocity of the main part of the stream observed by S15, they may not be unusual for the fanned-out part of the stream. As Figure 4 of \citet{pea15} shows, the amplitude of variations in $v_{\rm gsr}$ in the fanned-out part of their simulated Pal 5 stream can be as high as 50 km~s$^{-1}$. The observed variation in the velocity of stars with $v_{\rm gsr} > 280$ km~s$^{-1}$ is not likely due to epicyclic motion of stars along tidal tails \citep{kue08}. For example, Figure 5 of \citet{kue15} shows that, at least in the case of the Pal 5 stream, this motion may induce variations in velocity of only $\sim5$ km~s$^{-1}$ at a given position along the stream.

By identifying these new members, we have extended the observed length of the Ophiuchus stream from 1.6 kpc (S15) to 3 kpc. The stream now extends to 6.3 kpc from the Sun (was 7.5 kpc in S15). If the easternmost star is not considered as a member (since it has the lowest velocity of the four stars, and the highest offset from the observed part of the stream), the observed length of the Ophiuchus stream becomes 2.2 kpc, and the the stream extends to 7 kpc from the Sun. While a longer stream implies a slightly longer time of disruption (now at $\sim400$ Myr), the tension with the age of the population (12 Gyr), and the orbit of the stream (radial period of 240 Myr and apocenter of 17 kpc) still remains.

We would like to note that to calculate the above length of the stream and its proximity to the Sun, we simply used \autoref{stream_dm} and galactic longitudes of new members. We used \autoref{stream_dm} to calculate the distances to new members because that is how they were initially selected for followup. They were selected because their colors and magnitudes are consistent with the color-magnitude diagram of the Ophiuchus stream at some galactic longitude $\ell$, and that diagram depends on \autoref{stream_dm}.

The tension between the length of the Ophiuchus stream and its orbit may be lessened now that we have detected fanning of the stream. When S15 concluded that the progenitor of the Ophiuchus stream could not have been on its current orbit (measured by S15) for more than $\sim400$ Myr, they assumed the stream is on a regular, non-resonant orbit. However, as we already mentioned in \autoref{introduction}, fanning can be a signature of a stream on a chaotic orbit \citep{far15,apw16}. If the fanning of the Ophiuchus stream is due to the stream being on a chaotic orbit, then the orbital parameters measured by S15 today were likely different in the past 12 Gyr. At least in principle, the progenitor could be on a chaotic orbit that allowed it to survive for $\sim11$ Gyr, only to be disrupted in the last 0.5 Gyr. Whether this is a likely scenario, is something that will require detailed modeling, and we leave it for a future publication.

\acknowledgments{B.S.~acknowledges funding from the European Research Council under the European Union’s Seventh Framework Programme (FP 7) ERC Grant Agreement n.~${\rm [321035]}$. A.P.W.~is supported by a National Science Foundation Graduate Research Fellowship under Grant No.~11-44155. The Pan-STARRS1 Surveys (PS1) have been made possible through contributions by the Institute for Astronomy, the University of Hawaii, the Pan-STARRS Project Office, the Max-Planck Society and its participating institutes, the Max Planck Institute for Astronomy, Heidelberg and the Max Planck Institute for Extraterrestrial Physics, Garching, The Johns Hopkins University, Durham University, the University of Edinburgh, the Queen's University Belfast, the Harvard-Smithsonian Center for Astrophysics, the Las Cumbres Observatory Global Telescope Network Incorporated, the National Central University of Taiwan, the Space Telescope Science Institute, and the National Aeronautics and Space Administration under Grant No.~NNX08AR22G issued through the Planetary Science Division of the NASA Science Mission Directorate, the National Science Foundation Grant No.~AST-1238877, the University of Maryland, Eotvos Lorand University (ELTE), and the Los Alamos National Laboratory. Some of the data presented herein were obtained at the W.M.~Keck Observatory, which is operated as a scientific partnership among the California Institute of Technology, the University of California and the National Aeronautics and Space Administration. The Observatory was made possible by the generous financial support of the W.M.~Keck Foundation. The authors wish to recognize and acknowledge the very significant cultural role and reverence that the summit of Mauna Kea has always had within the indigenous Hawaiian community. Based on observations collected at the German-Spanish Astronomical Center, Calar Alto, jointly operated by the Max-Planck-Institut f\"{u}r Astronomie Heidelberg and the Instituto de Astrof\'{i}sica de Andalucía (CSIC).
}

{\it Facilities:} \facility{PS1}, \facility{Keck:I (DEIMOS)}, \facility{CAO:3.5m (TWIN)}

\bibliography{main}

\end{document}